# Distributed Stabilization by Probability Control for Deterministic-Stochastic Large Scale Systems: Dissipativity Approach

Koji Tsumura, Binh Minh Nguyen, Hisaya Wakayama, and Shinji Hara

*Abstract* – **By using dissipativity approach, we establish the stability condition for the feedback connection of a deterministic dynamical system Σ and a stochastic memoryless map Ψ. After that, we extend the result to the class of large scale systems in which: Σ consists of many sub-systems; and Ψ consists of many "stochastic actuators" and "probability controllers" that control the actuator's output events. We will demonstrate the proposed approach by showing the design procedures to globally stabilize the manufacturing systems while locally balance the stock levels in any production process.**

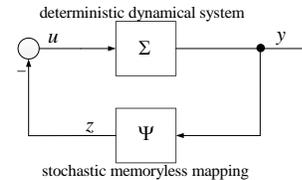

Fig. 1. Connection of deterministic system and stochastic map.

## I. Introduction

In these decades, one of the most active research fields in the control community is the distributed/decentralized control of large-scale systems, and it is also promoted by several research projects such as "Industry 4.0", "Smart factory," or "IoT." Our research group also have contributed to this field with a concept of "glocal control" [1] such as "generaized frequency variable" [2], "hierarchical decentralized LQR" [3], "control of power networks" [4], "in-wheel-motor vehicle" [5] and so on.

In order to make the availability of the distributed control strategy to the actual systems higher, it is necessary to extend the class of objective plants or controllers close to realistic systems. When the number of subsystems becomes large, it is natural that the properties and the behavior emerged in the systems become heterogeneous and from above idea, we consider large scale hybrid systems composed of deterministic subsystems, which may include time delay, and stochastic subsystems in this paper. In particular, we deal with a case that the actions of control actuators are stochastic.

We can find this class of systems in various applications such as electric power networks, supply chain systems, inventory management systems, manufacturing factories, chemical process systems, and biological systems, and so on.

A typical example among them is the power network with demand response performed by heating and refrigerator devices [6], [7]. The devices are usually distributed geographically and the action of the status ON or OFF of the devices can be modeled as stochastic behavior. Although the advantage of demand response for managing such behaviour has been evaluated by simulations in many works including [6] and [7], it is required to establish a theoretical framework to stabilize such power network with demand response.

Another example is a case of manufacturing factories, which are composed of several manufacturing processes and transportation processes of materials by a swarm of vehicles. One of strategies to operate the transportation is a centralized deterministic scheduling of all vehicles, however when the number of vehicles is large, it is not feasible. Another strategy is that each vehicle moves around autonomously and randomly according to guidance signals. This method is more realistic for large scale systems as discussed in this paper, however the behavior of the vehicles can be regarded as stochastic actions. As a result, the whole system is composed of deterministic manufacturing processes and stochastic transportation processes. Similar applications to this model include supply chains or inventory management systems.

Finally, chemical process control [8] and biological control system [9] are also examples of hybrid of deterministic and stochastic systems.

An idea of this paper to operate such hybrid systems is to feedback-control the probabilities of the stochastic actions of the actuators. We call this as "probability control," that is, to control the probabilities. However, the following simple questions arise in such control systems; (i) what control law of the probability control is reasonable? (ii) can we show the stability of the whole system? In particular, the feedback control law of the probability control becomes an operator from the state variables or the outputs to the probability from 0 to 1 and it is inevitably nonlinear. In order to solve this difficulty, we consider to employ a notion of stochastic dissipativity. This idea can be explained with Fig. 1, where Σ is a deterministic plant while Ψ is a stochastic control operator. Stochastic dissipativity/passivity has been also studied [10], [11], [12], [13], however, almost the previous works consider cases of exogenous Gaussian noises, input/output delay, or the Markovian switching of the systems and our case has not been enough investigated.

A related result for the case of manufacturing factories is Kosmatopoulos *et al.* [14], however, it considers a case of completely deterministic systems. Cases of supply chains and inventory management systems are also studied in [15], [16], and [17], however they are completely deterministic. On the other hand, our proposed control strategy is superficially similar to consensus over random networks [18]-[21]. However, in [18]-[21], the probabilities of the random communications in the networks are fixed. On the other hand, our method is to control the probabilities of control actions.

The reminder of this paper is organized as follows. In Section II, a manufacturing factory is introduced as a motivating example. In Section III, we introduce the dissipativity for stochastic systems and the related results which are utilized in the following sections. In Section IV, we propose control strategies for two cases of the manufacturing factories without and with transportation delay and shown main results on the stability. In Section V, we evaluate the effectiveness of our proposed methods by numerical simulations, and conclude in Section VI.

## II. MOTIVATING EXAMPLE

In this section, we explain the class of large scale control systems discussed in this paper by introducing a motivating example, a manufacturing system.

### A. Outline of the manufacturing system

As shown in Fig. 2, we consider a network of $N$ production processes (PPs). Each production process PP $i$ includes a pair of an input buffer and an output buffer of materials where $x_I^i(t)$ and $x_O^i(t)$ represent their stock levels, respectively. Let $p^i(t)$ denote the production rate from the input-buffer to the output-buffer of the PP $i$, $u_a^{i \to j}(t)$ the transportation rate from the output-buffer of PP $i$ to the input-buffer of PP $j$, $\underset{\to}{M}^i$ the set of indices of PP from which materials are sent to input-buffer of PP $i$, and $\overset{\leftarrow}{M}^i$ the set of indices to which materials are sent from output-buffer of PP $i$. The control objectives are (i) to satisfy a given throughput of materials from the entrance to the exit of the manufacturing system, (ii) to keep all the stock levels of materials in the buffers equal in a normalized unit. Note that the objective (ii) is for avoiding overflow or shortage of materials in all the buffers.

### B. Deterministic dynamics of the stock level in PP

The stock level dynamics are given as follows
$$x_I^i(t+1) = x_I^i(t) - p^i(t) + \sum_{k \in \underset{\to}{M}^i} u_a^{k \to i}(t) \tag{1}$$

$$x_O^i(t+1) = x_O^i(t) + p^i(t) - \sum_{k \in \overset{\leftarrow}{M}^i} u_a^{i \to k}(t) \tag{2}$$

To achieve the objective (ii) in the local PP $i$, the local production rate is given as
$$p^i(t) = -h^i \left( x_O^i(t) - x_I^i(t) \right) \tag{3}$$
where $h^i$ is a production control gain which is selected between 0 and 1 to stabilize the local agent PP $i$. Define

$$x^i(t) = \begin{bmatrix} x_I^i(t) & x_O^i(t) \end{bmatrix}^T, u^i(t) = \begin{bmatrix} \sum_{k \in \underset{\to}{M}^i} u_a^{k \to i}(t) & -\sum_{k \in \overset{\leftarrow}{M}^i} u_a^{i \to k}(t) \end{bmatrix}^T \tag{4}$$

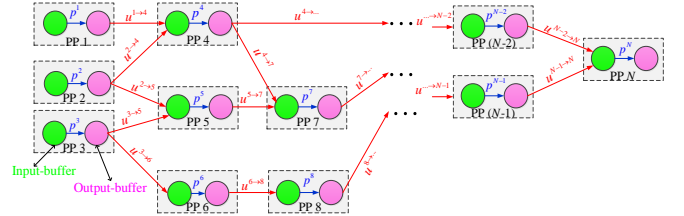

Fig. 2. Example of production flow in manufacturing systems.

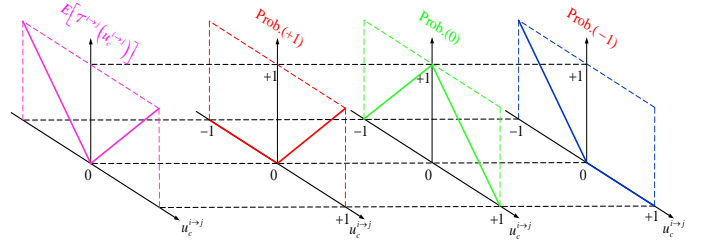

Fig. 3. Stochastic characteristics of the transportation actuator.

the stock level dynamics of the PP $i$ is expressed as
$$\mathcal{P}^i : x^i(t+1) = A^i x^i(t) + u^i(t) \tag{5}$$

$$A^i = \begin{bmatrix} 1-h^i & h^i \\ h^i & 1-h^i \end{bmatrix} \tag{6}$$

From (7), each $\mathcal{P}^i$ is shown to be a deterministic dynamical system. The stock levels $x^i$ can be measured by local sensor $\mathcal{S}_x^i$:
$$x_s^i(t) = \mathcal{S}_x^i \left( x^i(t) \right) = x^i(t) \tag{7}$$

### C. Stochastic transportation actuator

We assume that swarms of vehicles transport material flexibly and randomly from the output-buffer of PP $i$ to the input-buffer of PP $j$. There exists $M$ local transportation actuators $\mathcal{T}^{i \to j}, j \in \overset{\leftarrow}{M}^i, i \in [1, N]$. Each actuator can be treated as a memoryless operator that probabilistically outputs the quantized values $u_a^{i \to j}(t) \in \{0, +1, -1\}$. The input to each $\mathcal{T}^{i \to j}$ is a scalar signal $u_c^{i \to j}(t)$ generated by the transportation controller $\mathcal{C}_\mathcal{T}^{i \to j}$. The quantity "1" means a unit of materials under normalization. We regard the value "−1" of negative transportation as a "*virtual action*" from the concept of dynamic balancing [22]. The stochastic operator is defined as
$$\mathcal{T}^{i \to j}(u_c^{i \to j}(t)) = \begin{cases} +1 & \text{Prob.} = \pi_+(u_c^{i \to j}(t)) \times u_c^{i \to j}(t) \\ 0 & \text{Prob.} = 1 - |u_c^{i \to j}(t)| \\ -1 & \text{Prob.} = \pi_-(u_c^{i \to j}(t)) \times u_c^{i \to j}(t) \end{cases} \tag{8}$$

where $u_c^{i \to j}(t)$ is normalized between −1 and +1, and

$$\pi_+(u_c^{i \to j}(t)) = \begin{cases} +1 & \text{if } u_c^{i \to j}(t) \geq 0 \\ 0 & \text{if } u_c^{i \to j}(t) < 0 \end{cases}, \; \pi_-(u_c^{i \to j}(t)) = \begin{cases} -1 & \text{if } u_c^{i \to j}(t) \leq 0 \\ 0 & \text{if } u_c^{i \to j}(t) > 0 \end{cases}$$

From (9), the expectation of the output of the actuator is
$$E\left[ \mathcal{T}^{i \to j}(u_c^{i \to j}(t)) \right] = \pi_+(u_c^{i \to j}(t)) \times u_c^{i \to j}(t) - \pi_-(u_c^{i \to j}(t)) \times u_c^{i \to j}(t) \;.$$
The stochastic actuator and the expectation calculation is described in Fig. 3 for an easy understanding.

### D. Model of the overall manufacturing system

Now we define

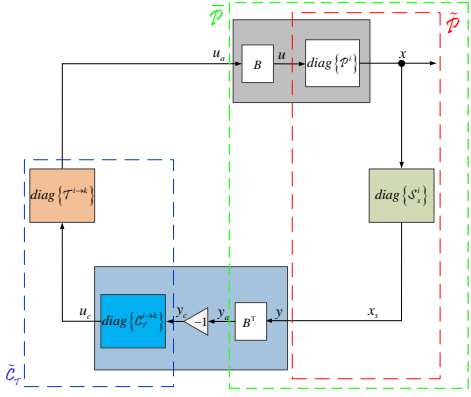

Fig. 4. Block diagram of manufacturing factory control system.

$$x(t) = [x^{1T}(t) \quad x^{2T}(t) \quad \ldots \quad x^{NT}(t)]^T \quad (9)$$

$$u_i(t) = [u_i^{1T}(t) \quad u_i^{2T}(t) \quad \ldots \quad u_i^{NT}(t)]^T \quad (10)$$

$$u_a(t) = [u_a^{1T}(t) \quad u_a^{2T}(t) \quad \ldots \quad u_a^{NT}(t)]^T \quad (11)$$

where the $i$th component $u_a^i(t) = \{u_a^{i \to k}(t)\}, k \in \overleftarrow{M}^i$, is a column vector of size $|\overleftarrow{M}^i|$. The dynamics of all stock levels is

$$x(t+1) = Ax(t) + Bu_a(t) = Ax(t) + u(t) \quad (12)$$

where $A = diag\{A^i\}$ and $B$ is an incidence matrix of the manufacturing network. The size of matrix $B$ is $2N \times M$ where $M = |\overleftarrow{M}^1| + \ldots + |\overleftarrow{M}^N|$. Due to space limitation, we neglect to discuss the general structure of matrix $B$ which was presented in our recent works [22]. For example, we consider the cyclic manufacturing system with the network structure described as: $\overleftarrow{M}^i = \{i+1\}$ for $i$ from 1 to $N-1$, and $\overleftarrow{M}^N = \{1\}$. $B_{cyclic} = [b_{ij}]$ where $0 \leq i \leq 2N$, $0 \leq j \leq N$. If $i = 1$ then the entry $b_{ij} = 1$ if $j = N$; otherwise $b_{ij} = 0$. If $2 \leq i \leq 2N$, then the entry $b_{ij} = (-1)^i$ if $i = 2j$ or $i = 2j+1$; otherwise $b_{ij} = 0$. We notice that

$$-B_{cyclic}^T x(t) = \begin{bmatrix} -x_I^2(t) + x_O^1(t) \\ -x_I^3(t) + x_O^2(t) \\ \vdots \\ -x_I^1(t) + x_O^N(t) \end{bmatrix} = \begin{bmatrix} y_c^{1 \to 2}(t) \\ y_c^{2 \to 3}(t) \\ \vdots \\ y_c^{N \to 1}(t) \end{bmatrix} \quad (13)$$

In general case, $y_c(t) = -B^T x(t)$ is a column vector that consists of the components $\{y_c^{i \to j}(t)\}$. It measures the stock level unbalances between the PPs. In other words, $y_c(t)$ can be utilized to attain the aforementioned global objective. It should be treated as the input to the decentralized transportation controller $\{C_\tau^{i \to j}\}$. Each $C_\tau^{i \to j}$ maps the input $y_c^{i \to j}(t)$ to the scalar control signal $u_c^{i \to j}(t)$ for controlling the local transportation actuator $\mathcal{T}^{i \to j}$. Based on (1)~(12), the overall manufacturing control system can be established as in Fig. 4.

*E. Disscusion*

Let $\tilde{C}_\tau$ be the cascade connection of the transportation controllers and the transportation actuators, and $\bar{\mathcal{P}}$ is the cascade connection of the stock level dynamics with the pair of matrices $\{B, B^T\}$. The system in Fig. 3 is equivalent to the feedback connection of a deterministic $\bar{\mathcal{P}}$ and a stochastic $\tilde{C}_\tau$. This feedback system model is not restricted to manufacturing systems, but other applications as mentioned in Section I. Stabilization of this system is a non-trivial question, since the system is quite complex and consists of many local sub-systems with different characteristics. The problem is even more complicated if the delay is introduced to the network. This motivates us to employ the dissipativity approach presented in the following section.

III. DETERMINISTIC-STOCHASTIC DISSIPATIVITY APPROACH

*A. Definitions and preliminary result*

Firstly, we consider a class of feedback connection shown in Fig. 1 where $\Sigma$ is a linear discrete-time system expressed as

$$\begin{cases} x(t+1) = Ax(t) + Bu(t) \\ y(t) = Cx(t) + Du(t) \end{cases} \quad (14)$$

where the input vector $u(t) \in \mathbb{R}^p$, the output vector $y(t) \in \mathbb{R}^p$, and the state vector $x(t) \in \mathbb{R}^n$; the matrices $\{A, B, C, D\}$ are of appropriate sizes. $\Psi$ is a stochastic memoryless map with the input vector $y(t)$ and the output vector $z(t)$ of the same size. This paper only considers the quadratic type of the supply rate, and the dissipativity of $\Sigma$ is stated as follows [23].

**Definition 1**: $\Sigma$ is $(Q_\Sigma, R_\Sigma, S_\Sigma)$ dissipative if there exists a positive semidefinite storage function $V: \mathbb{R}^n \to \mathbb{R}_+$, such that

$$V(x(t+1)) - V(x(t)) \leq W_\Sigma(y(t), u(t))$$

holds true $\forall x(t) \in \mathbb{R}^n$ and $u(t) \in \mathbb{R}^p$ where

$$W_\Sigma(y(t), u(t)) = y^T(t)Q_\Sigma y(t) + u^T(t)R_\Sigma u(t) + 2y^T(t)S_\Sigma u(t) \quad (15)$$

where $Q_\Sigma = Q_\Sigma^T, R_\Sigma = R_\Sigma^T$, and $S_\Sigma$ are the matrices of appropriate sizes.

**Definition 2**: We consider a map $g(\cdot)$ that maps an input $\alpha \in \mathbb{R}^p$ to the limited set $\{\beta^1, \beta^2, \ldots, \beta^K\}, \beta^k \in \mathbb{R}^p$. We let W be a function of $\alpha$ and $g(\alpha)$, W: $\mathbb{R}^p \times \mathbb{R}^p \to \mathbb{R}$. $W(\alpha, g(\alpha))$ might take one value in the limited set of $\{W(\alpha, \beta^k)\}$ with the probability $\{P^k(\alpha)\}$ where $k \in [1, K]$. $P^k(\alpha)$ is obtained by the stochastic properties of the map $g(\cdot)$ and it s.t. $P^1(\alpha) + \ldots + P^K(\alpha) = 1$, and $0 \leq P^k(\alpha) \leq 1$ for all $k$. The expectation of $W(\alpha, g(\alpha))$ is given as

$$E[W(\alpha, g(\alpha))] := \sum_{k=1}^{K} W(\alpha, \beta^k) P^k(\alpha) \quad (16)$$

**Definition 3**: $\Psi$ is said to be $(Q_\Psi, R_\Psi, S_\Psi)$ dissipative if the following inequality holds true

$$\sum_{i=0}^{l} E[W_\Psi(z(i), y(i))] \geq 0 \quad \forall y(i) \in \mathbb{R}^p, l \in \mathbb{Z}, l \geq 0$$

where

$$W_\Psi(z(t), y(t)) = z^T(t)Q_\Psi z(t) + y^T(t)R_\Psi y(t) + 2z^T(t)S_\Psi y(t) \quad (17)$$

where $Q_\Psi = Q_\Psi^T, R_\Psi = R_\Psi^T$, and $S_\Psi$ are the matrices of appropriate sizes.

For instance, $\Psi$ is said to be output strictly passive (OSP) if it is $(\Theta_p, -\delta I_p, 0.5I_p)$ for some scalar $\delta \geq 0$. Here $\Theta_p$ is the zero matrix, $I_p$ is the unity matrix. In this study, we are interested in the class of OSP $\Psi$ such that the supply rate satisfies the following condition for all time step $t$

$$E[z^T(t)y(t)] \geq \delta E[z^T(t)z(t)] \quad (18)$$

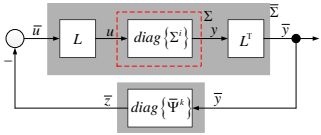

Fig. 5. Connection of multi deterministic systems and multi stochastic maps.

In (18), the expectation on both sides can also be calculated using **Definition 1**. Due to the stochastic properties of $\Psi$, the scalars $z^T(t)y(t)$ and $z^T(t)z(t)$ may take one of the values in the limited set $\{zy^k\}$ and $\{zz^k\}$, repsectively. The probabilities that $z^T(t)y(t) = zy^k$ and $z^T(t)z(t) = zz^k$ are assumed to be obtained from the stochastic properties of $\Psi$ with a given $y(t)$.

**Proposition 4**: If $\Sigma$ is ($\Theta_p$, $\gamma I_p$, $0.5I_p$,) dissipative with a storage function $V$, and $\Psi$ satisfies (18) for some scalar $\delta \geq \gamma$, then the following inequality holds true:

$$E\left[V\left(x(t+1)\right)\big|x(t)\right] - V\left(x(t)\right) \leq 0 \quad (19)$$

**Proof**: Since $\Sigma$ is ($\Theta_p$, $\gamma I_p$, $0.5I_p$,) dissipative and $u(t) = -z(t)$, the following inequality holds true

$$V\left(x(t+1)\right) - V\left(x(t)\right) \leq -\left(z^T(t)y(t) - \gamma z^T(t)z(t)\right) \quad (20)$$

As $\delta \geq \gamma$, the following inequality is established

$$V\left(x(t+1)\right) - V\left(x(t)\right) \leq -\left(z^T(t)y(t) - \delta z^T(t)z(t)\right) \quad (21)$$

Taking the expectations of the both sides of (21) with respect to (18), we finally obtain the inequality (19).

### B. Dissipativity for multi-agent systems

In this sub-section, we will extend **Proposition 4** to the class of multi-agent systems in Fig. 5. Each local system $\Sigma^i$ ($i$ from 1 to $N$) is a linear discrete time system with the input $u^i(t) \in \mathbb{R}^p$, output $y^i(t) \in \mathbb{R}^p$, and the state $x^i(t) \in \mathbb{R}^n$. Each $\bar{\Psi}^k$ ($k$ from 1 to $M$) is a stochastic map with the scalar input $\bar{y}^k(t)$ and the scalar output $\bar{z}^k(t)$. Matrix $L$ of size $Np \times M$ represents the connections between the agents. As discussed in Section II, matrix $L^T$ can be used for the purpose of consensus and system stabilization. The system $\Sigma = diag\{\Sigma^i\}$ has the vectors

$$u(t) = [u^{1T}(t) \quad u^{2T}(t) \quad \ldots \quad u^{NT}(t)]^T \in \mathbb{R}^{Np} \quad (22)$$
$$y(t) = [y^{1T}(t) \quad y^{2T}(t) \quad \ldots \quad y^{NT}(t)]^T \in \mathbb{R}^{Np} \quad (23)$$
$$x(t) = [x^{1T}(t) \quad x^{2T}(t) \quad \ldots \quad x^{NT}(t)]^T \in \mathbb{R}^{Nn} \quad (24)$$

The system $\bar{\Sigma}$ has the input $\bar{u}(t)$ and the output $\bar{y}(t)$. The map $\bar{\Psi} = diag\{\bar{\Psi}^i\}$ has the input vector $\bar{y}(t)$ and the output vector $\bar{z}(t)$ expressed as

$$\bar{y}(t) = \begin{bmatrix} \bar{y}^1(t) & \bar{y}^2(t) & \ldots & \bar{y}^M(t) \end{bmatrix}^T \in \mathbb{R}^M \quad (25)$$
$$\bar{z}(t) = \begin{bmatrix} \bar{z}^1(t) & \bar{z}^2(t) & \ldots & \bar{z}^M(t) \end{bmatrix}^T \in \mathbb{R}^M \quad (26)$$

**Proposition 5**: In Fig. 5, we assume that each agent $\Sigma^i$ is ($\Theta_p, \gamma^i I_p, 0.5 I_p$) dissipative with a storage function $V^i$; and each $\bar{\Psi}^k$ is an OSP map that s.t. $E[\bar{z}^k(t)\bar{y}^k(t)] \geq \delta^i E[(\bar{z}^k(t))^2]$.

If $diag\{\delta^k\} \geq L^T(diag\{\gamma^i I_p\})L$, then there exists a storage function $V$ such that the following inequality holds true $E[V(x(t+1))|x(t)] - V(x(t)) \leq 0$.

**Proof**: Since each $\Sigma^i$ is ($\Theta_p, \gamma^i I_p, 0.5 I_p$) dissipative, the following inequality holds true for the system $\Sigma$:

$$V\left(x(t+1)\right) - V\left(x(t)\right) \leq \left(y(t) + (diag\{\gamma^i I_p\})u(t)\right)^T u(t) \quad (27)$$

where the storage function $V = V^1 + V^2 + \ldots + V^N$. Substitute $u(t) = L\bar{u}(t)$ and $\bar{y}(t) = L^T y(t)$ into (27), we have

$$V(x(t+1)) - V(x(t)) \leq -\bar{z}^T(t)\bar{y}(t) + \bar{z}^T(t)L^T(diag\{\gamma^i I_p\})L\bar{z}(t) \quad (28)$$

If $\{\delta^k\}$ s.t. $diag\{\delta^k\} \geq L^T(diag\{\gamma^i I_p\})L$, we can obtain the following inequality with respect to the OSP of each $\bar{\Psi}^k$:

$$E[\bar{z}^T(t)\bar{y}(t)] \geq E[\bar{z}^T(t)L^T(diag\{\gamma^i I_p\})L\bar{z}(t)] \quad (29)$$

Taking the expectations of the left and right hand sides of (28) with respect to (29), it is finally shown that the inequality $E[V(x(t+1))|x(t)] - V(x(t)) \leq 0$ holds true.

## IV. DEMONSTRATION OF THE PROPOSED DISSIPATIVITY APPROACH

This Section is to demonstrate the dissipativity approach presented in Section III. We consider the cyclic manufacturing systems with homogeneous production control gains just for simplicity. We will examine two cases, without and with transportation delay, respectively.

### A. Cyclic manufacturing system without transportation delay

Main result:

The procedure to design the transportation controller is proposed as follows.

**Procedure 1 (without transportation delay)**

We consider a cyclic manufacturing system of $N$ PPs with the homogeneous production control gain $h^i = h$ for all $i$ from 1 to $N$. Let $\gamma_m = 1/(2(1-h))$. The transportation controller $\mathcal{C}_\tau^{i \to j}$ is designed such that

(i) It is a SISO memoryless mapping with the input $y_c^{i \to j}(t)$ and the output $u_c^{i \to j}(t)$ which is a normalized number that varies between $-1$ and $+1$.

(ii) The cascade connection of $\mathcal{C}_\tau^{i \to j}$ and $\mathcal{T}^{i \to j}$ is a memoryless map $\tilde{\mathcal{C}}_\tau^{i \to j}$ with the input $y_c^{i \to j}(t)$ and the output $u_a^{i \to j}(t)$ that satisfies $E[u_a^{i \to j}(t) y_c^{i \to j}(t)] \geq \delta E[(u_a^{i \to j}(t))^2]$ where $\delta \geq 2\gamma_m$.

Stability and consensus of the overall system:

**Proposition 6**: If the cyclic manufacturing system is design to satisfy the **Procedure 1**, then the following inequality holds true for the total network $E[V(x(t+1))|x(t)] - V(x(t)) \leq 0$ where $V(x(t)) = 0.5 x^T(t) x(t)$.

**Proof**: A sketch of the proof is presented as follows. Let $\tilde{\mathcal{P}}^i$ be the cascade connection of $\mathcal{P}^i$ and $\mathcal{S}_x^i$, with the input $u^i(t)$ and the output $y^i(t) = x_s^i(t)$. Applying the KYP Lemma [27], each $\tilde{\mathcal{P}}^i$ is shown to be ($\Theta_2, \gamma I_2, 0.5 I_2$) dissipative with the storage function $V^i(x^i(t)) = 0.5 x^{iT}(t) x^i(t)$ and $\gamma \geq \gamma_m = 1/(2(1-h))$.

We apply **Proposition 5** to the system in Fig. 4 in which $N$ agents $\{\tilde{\mathcal{P}}^i\}$ interconnected with $N$ memoryless maps $\{\tilde{\mathcal{C}}_\tau^{i \to k}\}$ via a pair of matrices $\{B, B^T\}$. The stability condition for mapping $\tilde{\mathcal{C}}_\tau^{i \to k}$ is $\delta I_N \geq \gamma B^T B$. We notice that $B$ is the incidence matrix, and graph theory shows that $B^T B = 2I_N + \Delta$ where $\Delta$ is the adjacency matrix of the manufacturing network's line graph [24]. In case of cyclic manufacturing, $\Delta$ is a zero matrix. Thus, the condition $\delta I_N \geq \gamma B^T B$ is equivalent to $\delta \geq 2\gamma$. By selecting $\delta \geq 2\gamma_m$, it is enough to guarantee that the overall manufacturing network s.t. $E[V(x(t+1))|x(t)] - V(x(t)) \leq 0$ where the storage function is $V(x(t)) = 0.5x^T(t)x(t)$. This completes the proof.

Next, we define the equilibrium state vector

$$\bar{x} = \left(\frac{1}{2N}\mathbf{1}_{2N}^T \cdot x(0)\right)\mathbf{1}_{2N}$$

where $\mathbf{1}_{2N}$ is the all-one column vector of size $2N$, and $x(0)$ is the initial state vector that includes all the stock levels at time step $t = 0$. The error state vector is defined as $\tilde{x}(t) = x(t) - \bar{x}$.

**Proposition 7**: If the cyclic manufacturing system designed to satisfy the **Procedure 1**, then the error state dynamics s.t.

$$E[V(\tilde{x}(t+1))|\tilde{x}(t)] - V(\tilde{x}(t)) \leq 0, \forall t \quad (30)$$

where $V(\tilde{x}(t)) = 0.5\tilde{x}^T(t)\tilde{x}(t)$.

**Proof**: The system the system $\bar{\mathcal{P}}$ in Fig. 4 is expressed as

$$\begin{cases} x(t+1) = Ax(t) + Bu_a(t) \\ y_a(t) = B^T x(t) \end{cases} \quad (31)$$

From the structure of the matrices $A$ and $B$ and the definition of the error state vector, we have

$$\begin{cases} \tilde{x}(t+1) = A\tilde{x}(t) + Bu_a(t) \\ y_a(t) = B^T \tilde{x}(t) \end{cases} \quad (32)$$

Then we can treat the overall system as the feedback connection of the error dynamical system (32) and the stochastic map $\tilde{\mathcal{C}}_\tau = diag\{\tilde{\mathcal{C}}_\tau^{i\to k}\}$. The error system is shown to be $(\Theta_N, 2\gamma_m I_N, 0.5I_N)$ dissipative with the storage function $V(\tilde{x}(t)) = 0.5\tilde{x}^T(t)\tilde{x}(t)$. Applying the **Proposition 4**, we can finally conclude that the inequality (30) holds true.

Structure of the transportation controller:

A class of transportation controller that satisfies the **Procedure 1** is proposed as in Fig. 6 where $L^{i\to j}$ is a positive control gain and the function $f^{i\to j}(\cdot)$ is a normalizer with the following characteristics. (i) $f^{i\to j}(\theta)$ is monotonic increasing for $|\theta| \geq L^{i\to j}\delta$; (ii) $f^{i\to j}(0) = 0$ for $|\theta| < L^{i\to j}\delta$; (iii) $|f^{i\to j}(\theta)| \leq 1$ for all $\theta \in \Re$. Several possible normalizer candidates are:

$$f^{i\to j}(\theta) = \begin{cases} \tanh(\theta) & \text{if } |\theta| \geq L^{i\to j}\delta \\ 0 & \text{if } |\theta| < L^{i\to j}\delta \end{cases} \quad (33\text{-}a)$$

$$f^{i\to j}(\theta) = \begin{cases} \frac{2}{\pi}\tan^{-1}(\theta) & \text{if } |\theta| \geq L^{i\to j}\delta \\ 0 & \text{if } |\theta| < L^{i\to j}\delta \end{cases} \quad (33\text{-}b)$$

$$f^{i\to j}(\theta) = \begin{cases} \frac{\theta}{\theta_m} & \text{if } L^{i\to j}\delta \leq |\theta| < \theta_m \\ 0 & \text{if } |\theta| < L^{i\to j}\delta \\ \text{sign}(\theta) & \text{if } |\theta| \geq \theta_m \end{cases} \quad (33\text{-}c)$$

where $\theta_m$ is a suitable threshold.

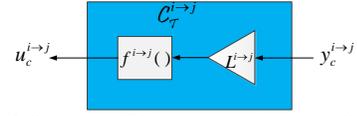

Fig. 6. Structure of the transportation controller.

### B. Cyclic manufacturing system with transportation delay

Model of the system with transportation delay:

We notice that the stock level in the output-buffer of PP $i$ can be reduced instantly. We assume that the transportation route to the input-buffer of PP $i$ has a delay of $d^i$ steps which is a positive integer known by the controller. Define the delay operator $\Gamma^i = diag\{\rho^i(\,), 1\}$ where where $\rho^i(\,)$ is the SISO operator that delays the scalar input by $d^i$ steps. The dynamics of the stock levels at PP $i$ can be written as

$$x^i(t+1) = A^i x^i(t) + \Gamma^i(u^i(t)) = A^i x^i(t) + B_O u^i(t) + B_I u^i(t-d^i) \quad (34)$$

$$B_I = \begin{bmatrix} 1 & 0 \\ 0 & 0 \end{bmatrix}, B_O = I_2 - B_I \quad (35)$$

It is well known that if the delay is not handled properly, the consensus performance and system stability would be degraded. For this reason, we will not use $y_c(t) = -B^T x(t)$ as the input to the transportation controller. Instead, we design for each local PP $i$ a "delay compensator" $\Xi^i$. Each $\Xi^i$ utilizes the local stock levels measurement at time $t$ and the input signals at several previous steps to calculate a delay-handled signal $y^i(t)$. The local delay compensation is described as in Fig. 7, and the overall manufacturing system is proposed as in Fig. 8. The main results are summarized as follows.

Proposal of delay compensator:

We can re-express the delay dynamics (34) as

$$\bar{x}^i(t+1) = \bar{A}^i \bar{x}^i(t) + \bar{B}^i u^i(t) \quad (36)$$

$$\bar{x}^i(t) = \begin{bmatrix} x^{iT}(t) & z_{d^i}^T(t) & z_{d^i-1}^T(t) & \cdots & z_2^T(t) & z_1^T(t) \end{bmatrix}^T \quad (37)$$

$$z_{d^i}(t) = u^i(t-d^i), z_{d^i-1}(t) = u(t-d^i+1),\ldots,z_1 = u(t-1) \quad (38)$$

$$\bar{A}^i = \begin{bmatrix} A^i & B_I & \Theta_2 & \cdots & \Theta_2 & \Theta_2 \\ \Theta_2 & \Theta_2 & I_2 & \cdots & \Theta_2 & \Theta_2 \\ \Theta_2 & \Theta_2 & \Theta_2 & \cdots & \Theta_2 & \Theta_2 \\ \vdots & \vdots & \vdots & \ddots & \ddots & \vdots \\ \Theta_2 & \Theta_2 & \Theta_2 & \cdots & \Theta_2 & I_2 \\ \Theta_2 & \Theta_2 & \Theta_2 & \cdots & \Theta_2 & \Theta_2 \end{bmatrix}, \bar{B}^i = \begin{bmatrix} B_O \\ \Theta_2 \\ \Theta_2 \\ \vdots \\ \Theta_2 \\ I_2 \end{bmatrix} \quad (39)$$

where $\Theta_2$ is the 2×2 zero matrix, and $I_2$ is the 2×2 unity matrix.

**Proposition 8**: For each PP $i$, the partitioned matrix $\Omega^i = \{\Omega_{j,k}^i\}$ is established by the following procedure

$$\Omega_{j,j}^i = I_2, j = 1,2,\ldots,d^i+1 \quad (40\text{-}a)$$

$$\Omega_{1,2}^i = A^i \Omega_{1,1}^i B_I, \Omega_{1,j}^i = A^i \Omega_{1,j-1}^i \; (j = 3,4,\ldots,d^i+1) \quad (40\text{-}b)$$

$$\Omega_{2,j}^i = B_I \Omega_{1,j-1}^i \; (j = 3,4,\ldots,d^i+1) \quad (40\text{-}c)$$

$$\Omega_{j,k}^i = \Omega_{j-1,k-1}^i \; (j = 3,4,\ldots,d^i+1 \,\&\, k = j+1,\ldots,d^i+1) \quad (40\text{-}d)$$

$$\Omega_{j,k}^i = \left(\Omega_{k,j}^i\right)^T \; (j = 2,4,\ldots,d^i+1 \,\&\, k = 1,\ldots,j-1) \quad (40\text{-}e)$$

Then, design for each PP $i$ a delay compensator as

$$\Xi^i : y^i(t) = C_1^i x_s(t) + \sum_{j=1}^{d^i} C_{j+1}^i u_s^i(t - d^i + j - 1) \quad (41\text{-}a)$$

$$C_1^i = B_O \Omega_{1,1}^i A^i + \Omega_{d^i+1,1}^i A^i \quad (41\text{-}b)$$

$$C_2^i = B_O \Omega_{1,1}^i B_I + \Omega_{d^i+1,1}^i B_I \quad (41\text{-}c)$$

$$C_j^i = B_O \Omega_{1,j-1}^i + \Omega_{d^i+1,j-1}^i B_I \quad (j = 3,4,\ldots d^i + 1) \quad (41\text{-}d)$$

Define

$$\overline{C}^i = \begin{bmatrix} C_1^i & C_2^i & \ldots & C_{d^i+1}^i \end{bmatrix}$$

The system $\tilde{\mathcal{P}}^i$ in Fig. 7 with the state space equation given as

$$\begin{cases} \overline{x}^i(t+1) = \overline{A}^i \overline{x}^i(t) + \overline{B}^i(t) u^i(t) \\ y^i(t) = \overline{C}^i \overline{x}^i(t) \end{cases} \quad (42)$$

is ($\Theta_2$, $\gamma^i I_2$, $0.5 I_2$) dissipative with the storage function

$$V^i(\overline{x}^i(t)) = 0.5 \overline{x}^{iT}(t) \Omega^i \overline{x}^i(t) \quad (43)$$

for all number $\gamma^i$ that satisfies

$$\gamma^i \geq \gamma_m^i = \frac{3 + \sqrt{1 + \left((1-2h)^{d^i} - 1\right)^2}}{4} \quad (44)$$

**Proof**: A sketch of the proof is explained as follows. Given a production control gain $h \in (0, 1)$, by applying Cholesky decomposition, matrix $\Omega^i$ is shown to be positive definite for all positive integer number of $d^i$. This means we can select the class of storage function (43) for the augmented system $\tilde{\mathcal{P}}^i$. Following the KYP Lemma, each $\tilde{\mathcal{P}}^i$ is ($\Theta_2$, $2\gamma^i I_2$, $0.5 I_2$) dissipative with the storage function (43) if the following matrix is negative semi-definite

$$\overline{Q}^i = \begin{bmatrix} \overline{A}^{iT} \Omega^i \overline{A}^i - \Omega^i & \overline{A}^{iT} \Omega^i \overline{B}^i - \overline{C}^{iT} \\ \overline{B}^{iT} \Omega^i \overline{A}^i - \overline{C}^i & \overline{B}^{iT} \Omega^i \overline{B}^i - 2\gamma^i I_2 \end{bmatrix} \quad (45)$$

From (39), (40-a,b,c,d,e) and (41-a,b,c,d), the components $\overline{A}^{iT} \Omega^i \overline{B}^i - \overline{C}^{iT}$ and $\overline{B}^{iT} \Omega^i \overline{A}^i - \overline{C}^i$ are shown to be zero matrices, and $\overline{A}^{iT} \Omega^i \overline{A}^i - \Omega^i$ is a negative semi-definite matrix. We only need to take care of the negative semi-definiteness of $\overline{B}^{iT} \Omega^i \overline{B}^i - 2\gamma^i I_2$ which finally lets us to (44).

Based on **Proposition 8**, we consider a cyclic manufacturing system with $h^i = h$ for all $i$ from 1 to $N$. Each PP $i$ has an input transportation delay of $d^i$ step which is a known integer.

**Procedure 2 (with constant transportation delay)**

Step 1: Design for each PP $i$ a delay compensator $\Xi^i$ by using (40-a,b,c,d,e) and (41-a,b,c,d).

Step 2: Design for each transportation route $j \to i$ a transportation controller $\mathcal{C}_\tau^{j \to i}$ such that:

(i) the output $u_c^{j \to i}(t)$ of $\mathcal{C}_\tau^{j \to i}$ is a normalized number that varies between -1 and +1.

(ii) the cascade connection $\tilde{\mathcal{C}}_\tau^{j \to i}$ of $\mathcal{C}_\tau^{j \to i}$ and $\mathcal{T}^{j \to i}$ with input $y_c^{j \to i}(t)$ and output $u_a^{j \to i}(t)$ is a memoryless map that satisfies

$$E\left[u_a^{j \to i}(t) y_c^{j \to i}(t)\right] \geq \delta^{j \to i} E\left[\left(u_a^{j \to i}(t)\right)^2\right] \quad (46)$$

where the set of $\{\delta^{j \to i}\}$ satisfies

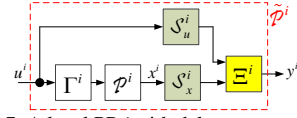

Fig. 7. A local PP $i$ with delay compensator.

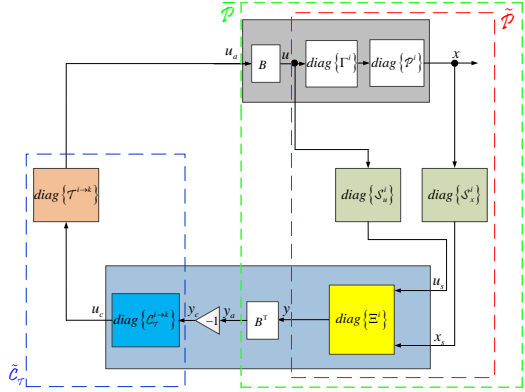

Fig. 8. Manufacturing factory control system with delay compensator.

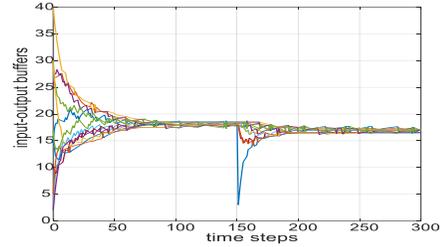

Fig. 9. Test 1-Case 1 (Without transportation delay).

$$\text{diag}\{\delta^{j \to i}\} \geq B^T \left(\text{diag}\{\gamma_m^i I_2\}\right) B \quad (47)$$

Stability and consensus of the overall system:

We neglect to show the proof of system stability which is very straightforward. To discuss the consensus performance, we define $\overline{x}^i = \left(\frac{1}{2N} \mathbf{1}_{2N}^T \cdot x(0)\right) \mathbf{1}_2$ and $\tilde{x}^i(t) = x^i(t) - \overline{x}^i$ where $\mathbf{1}_2 = [1\ 1]^T$. Then, we construct the augmented state

$$\overline{\tilde{x}}^i(t) = \begin{bmatrix} \tilde{x}^{iT}(t) & z_{d^i}^T(t) & z_{d^i-1}^T(t) & \ldots & z_2^T(t) & z_1^T(t) \end{bmatrix}^T \quad (48)$$

The following statement is obtained based on **Proposition 5**.

**Proposition 9**: If the cyclic manufacturing system is designed to satisfy the **Procedure 2**, then the following inequality holds true

$$E\left[V\left(\overline{\tilde{x}}(t+1)\right) \mid \overline{\tilde{x}}(t)\right] - V\left(\overline{\tilde{x}}(t)\right) \leq 0 \quad (49)$$

where $V = V^1 + V^2 + \ldots + V^N$ with $V^i\left(\overline{x}^i(t)\right) = 0.5 \overline{x}^{iT}(t) \Omega^i \overline{x}^i(t)$ and $\overline{\tilde{x}}(t) = \begin{bmatrix} \overline{\tilde{x}}^{1T}(t) & \overline{\tilde{x}}^{2T}(t) & \ldots & \overline{\tilde{x}}^{NT}(t) \end{bmatrix}^T$.

## V. SIMULATION VERIFICATION

### A. Outline of the simulation

We consider the cyclic manufacturing factory with 6 PPs. The simulation is performed from the initial until $t = 300$. The initial values of the input-buffers and output-buffers are {15, 40, 30, 10, 5, 2} and {27, 25, 2, 15, 30, 17}, respectively. We assume that at the input-buffer of PP 1, there is a rate ($\xi_I = 0.05$) of raw materials from the outside of the factory. We also assume

Table 1: Summary of the simulation setting

| Test | Case | Delay compensator | Transportation system $\tilde{\mathcal{C}}_\tau$ | $L^{i \to j}$ | Normalizer | $h^i$ | Design procedure |
|---|---|---|---|---|---|---|---|
| 1.Without delay | 1 | No | Strictly passive | 0.30 | (33-b) | 0.10 | **Procedure 1** |
| 2.Small delay ($d^i = 5$) | 2-1 | No | Passive | 0.75 | $f^{i \to k}(\theta) = (2/\pi)\tan^{-1}(\theta)$ | 0.10 | Consensus algorithm [22] |
| | 2-2 | No | Strictly passive | 0.75 | (33-b) | 0.10 | **Procedure 1** |
| | 2-3 | Yes, nominal delay $d_n = d^i$ | Strictly passive | 0.75 | (33-b) | 0.10 | **Procedure 2** |
| 3. Delay uncertainty ($d^i \in [8, 12]$) | 3-1 | No | Passive | 0.75 | $f^{i \to k}(\theta) = (2/\pi)\tan^{-1}(\theta)$ | 0.10 | Consensus algorithm [22] |
| | 3-2 | No | Strictly passive | 0.75 | (33-b) | 0.10 | **Procedure 1** |
| | 3-3 | Yes, nominal delay $d_n = \max\{d^i\}$ | Strictly passive | 0.75 | (33-b) | 0.10 | **Procedure 2** |

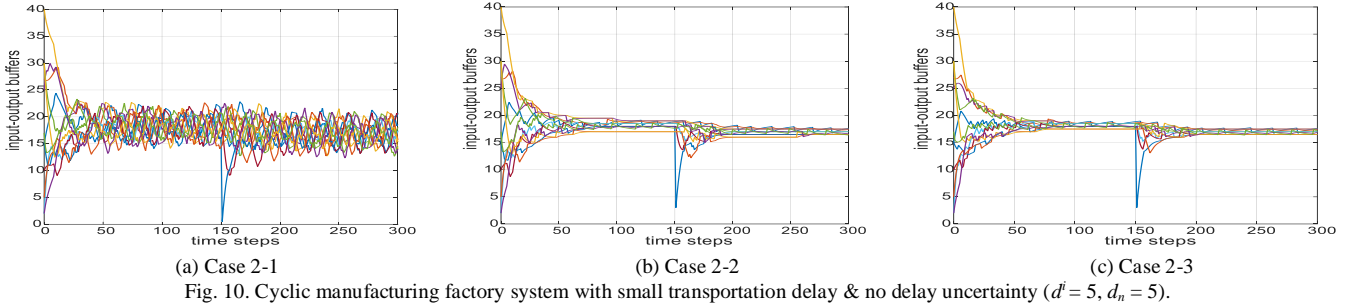

(a) Case 2-1  (b) Case 2-2  (c) Case 2-3
Fig. 10. Cyclic manufacturing factory system with small transportation delay & no delay uncertainty ($d^i = 5$, $d_n = 5$).

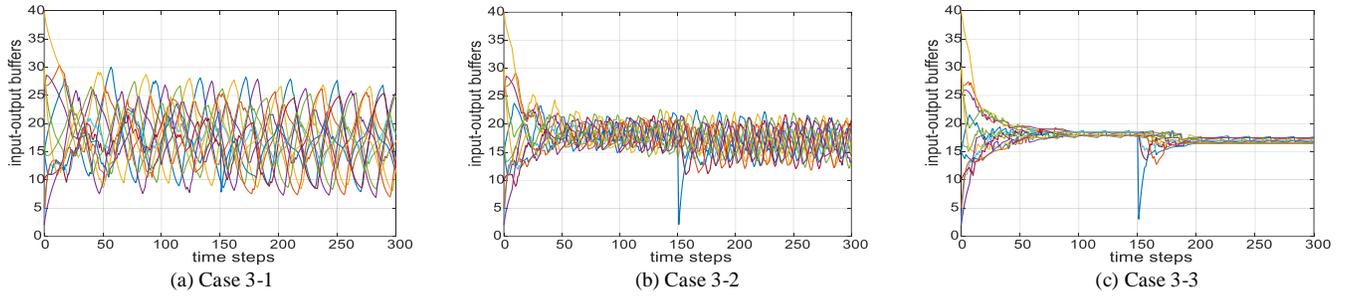

(a) Case 3-1  (b) Case 3-2  (c) Case 3-3
Fig. 11. Cyclic manufacturing factory system with large transportation delay & delay uncertainty ($d^i \in [8, 12]$, $d_n = \max\{d^i\}$).

there is a constant output rate ($\xi_O = 0.05$) of the produced production at the output-buffer of the PP 6. To evaluate the performance of the overall system under disturbance, we assume that at time step $t = 150$, a large quantity ($\zeta = 15$) of materials are taken out from the output-buffer of the PP 4. We conduct three simulation tests, and their setting conditions are summarized in Table 1.

### B. Test1: Without transportation delay (Fig. 9)

This test is to very the **Procedure 1**. As can be seen in Fig. 9, nice consensus performance is achieved by the proposed dissipativity approach. After about 50 steps, all stock levels converge to an equilibrium value which equals to the average of the initial values. At $t = 150$, the output-buffer of PP 4 is reduced a large amount of 15 units. Even though, the stability of the manufacturing system is guaranteed. A new consensus stage is re-established at time step $t \sim 200$. The results show that both local objective and global objective are successfully attained by the proposed approach.

### C. Test2: Small delay & no delay uncertainty (Fig. 10)

In this test, the delays in all transportation routes are $d^i = 5$ steps. The transportation control gain is set as 0.75 for clearly observing the influence of transportation delay. The simulation results are summarized in Fig. 10 as follows.

Case 2-1 (Fig. 10(a)): We use the conventional consensus algorithm proposed in our recent works [22]. The signal sent to the transportation controller is $y_c(t) = -B^T x_s(t)$. We select the arctangent normalizer as shown in Table 1. We can verify that $\tilde{\mathcal{C}}_\tau$ is only passive by this setting. Since the delay compensator is not implemented in this case, the stock levels suffer fluctuation of large amplitudes.

Case 2-2 (Fig. 10(b)): We use the proposed **Procedure 1**. In this test, the consensus stage is still maintained. This means the dissipativity approach can tolerate the transportation delay to a certain extent, even if the compensator is not implemented.

Case 2-3 (Fig. 10(c)): We use the **Procedure 2**. The delay compensator is implemented using the correct nominal delay $d_n = d^i = 5$. Thanks to the delay compensator, the consensus performance of Case 2-3 is much better than Case 2-1, and slightly improved in comparison with Case 2-2.

### D. Test 2: Large delay & delay uncertainty (Fig. 11)

In this test, the delay is quite large. We set $d^i$ as a number between 8 and 12. The simulation results of this test are summarized in Fig. 11 as follows.

Case 3-1 (Fig. 11(a)): We still use the consensus algorithm presented in [22]. The setting is similar to Case 2-1. It is transparent that the fluctuations of stock levels become very serious in this case.

Case 3-2 (Fig. 11(b)): The setting is similar to Case 2-2. The control performance is still better than Case 2-1, but the fluctuation become noticeable. This result shows that it is essential to handle the delay properly.

Case 3-3 (Fig. 11(c)): The system is designed using the **Procedure 2**. The controller does not know the exact delay of each transportation route, but its lower-bound and upper-bound. We design the delay compensator with the nominal delay $d_n = \max\{d^i\} = 12$. This means the control system suffers a delay uncertainty $\Delta d^i = d^i - d_n$ for each transportaion route. Even though, the simulation result shows that the consensus performance is still successfully attained. The consensus performance of Case 3-3 is almost comparable to Case 2-3 (small delay & no delay uncertainty) and Case 1 (without any delay). This means the delay compensator might tolerate the delay uncertainty to a certain extent. Our hypothesis is as follows: The overall system is still stable if the nominal delay is selected as the upper-bound of the actual delay. In future, we will investigate this hypothesis to explain it theoretically.

## VI. Conclusions

Based on dissipativity, this paper establishes the stability condition for a class of large scale discrete-time systems which consists of deterministic dynamical sub-systems and stochastic memoryless mapping. Our proposal is demonstrated by manufacturing systems as a case of study. The global and local objectives are attained in our dissipativity frameworks by only using the local measurements and local control actions. This paper, therefore, is an application of our glocal concept to the multi-agent systems. Moreover, the delay of material transportation between the agents can be handled properly by using the "delay compensator" designed based on dissipativity approach. The effectiveness of the proposed manufacturing control system is verified by numerical simulations.

In future study, we are interested in a theoretical explanation for the robustness of the proposed control system when it has to suffer delay uncertainty. We will also consider the situation such that the time delay is randomly changed in time. Besides that, we will examine the stochastic mapping with hierarchical decentralized structure for addressing the additional control objectives.